\documentclass[runningheads]{llncs}
\usepackage{graphicx}
\usepackage{comment}
\usepackage{parskip}

\begin{document}
\title{LinkRing:  A Wearable  Haptic  Display  for Delivering Multi-contact and Multi-modal Stimuli at the Finger Pads}
\titlerunning{A Wearable  Haptic  Display LinkRing}
%
\author{Aysien Ivanov*\inst{1} \and Daria Trinitatova*\inst{1} \and Dzmitry Tsetserukou\inst{1}
}
\authorrunning{A. Ivanov et al.}
%
\institute{Skolkovo Institute of Science and Technology (Skoltech),  Moscow 121205, Russia
\email{\{aysien.ivanov,daria.trinitatova,d.tsetserukou\}@skoltech.ru}\\
$\ast$Both authors contributed equally to the paper
}

\maketitle              
\begin{abstract}
LinkRing is a novel wearable tactile display for providing multi-contact and multi-modal stimuli at the finger. The system of two five-bar linkage mechanisms is designed to operate with two independent contact points, which combined can provide such stimulation as shear force and twist stimuli, slippage, and pressure. The proposed display has a lightweight and easy to wear structure. Two experiments were carried out in order to determine the sensitivity of the finger surface, the first one aimed to determine the location of the contact points, and the other for discrimination the slippage with varying rates. The results of the experiments showed a high level of pattern recognition. 

\keywords{Haptic display  \and Wearable tactile devices \and  Multi-contact stimuli \and Five-bar linkage}
\end{abstract}
\section{Introduction}

\setlength{\parindent}{3.5ex}Recent developments in computer graphics and head-mounted displays contributed to the appearance of different applications and games that implement VR technologies. Nevertheless, there is still a lack of providing haptic feedback, which is a crucial component to accomplish the full user immersion in a virtual environment.

Tactile information obtained from the fingers is one of the most important tools for interacting with the environment as fingertips have a rich set of mechanoreceptors. Thereby, the finger area plays an essential role in haptic perception. Nowadays, there are many research projects aimed at the design and application of fingertip haptic devices. One of the common methods for providing cutaneous feedback to the fingertips is by a moving platform shifting on the finger pad \cite{gabardi2016new}, \cite{chinello2015design}. Several works are focused at delivering mechanical stimulations and can reliably generate normal forces to the fingertip \cite{scheggi2015touch}, as well as shear deformation of the skin \cite{schorr2017three}. The most general method to simulate the object texture is using vibration \cite{Yatani2009}, \cite{Romano2010}, and the friction sensation is usually generated by electrostatic force  \cite{Meyer2014}. A  number of studies investigate the perception of object softness \cite{frediani2014wearable}, \cite{Yem2018}. Villa Salazar et al. \cite{VillaSalazar2020} examined the impact of combining simple passive tangible VR objects and wearable haptic display on haptic sensation by modifying the stiffness, friction, and shape perception of tangible objects in VR. Some research works explore multimodal tactile stimulation. In \cite{Yem2017}, Yem et al. presented FinGAR haptic device, which combines electrical and mechanical stimulation for generating skin deformation,  high/low-frequency vibration, and pressure. Similarly, the work \cite{Yem2019} presents the fingertip haptic interface for providing electrical, thermal, and vibrotactile stimulation.
Still, there is a need for a device that allows users to interact with virtual objects in a more realistic way. The interaction can be improved by the physical perception of the objects and their dimensions. Our approach suggests using a wearable haptic device, which can provide a multi-contact interaction on the finger pad that a person can experience when interacting with a real object.

\begin{figure}[h]
\centering
\vspace{-0.5em}
\includegraphics[width=\linewidth]{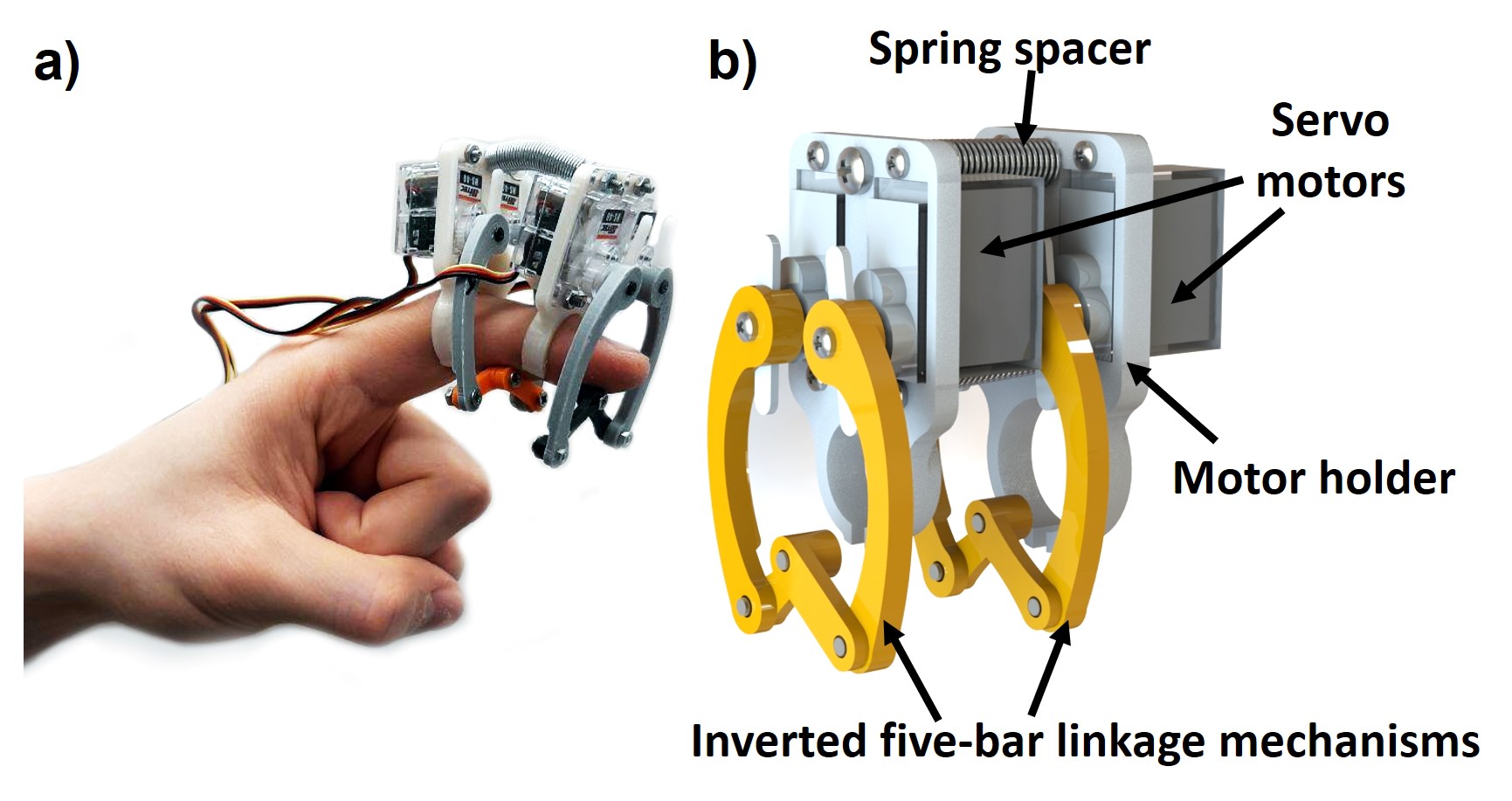}
\qquad
\vspace{-1.5em}
\caption{a) A wearable haptic display LinkRing. b) A CAD model of wearable tactile display LinkRing.}
\label{fig:main}
\vspace{-1.5em}
\end{figure}

In this paper, we propose LinkRing, a  novel wearable haptic device with 4-DoF, which provides multi-contact and multi-modal cutaneous feedback on the finger pad (Fig. \ref{fig:main}(a)). The haptic display is designed as the system of two five-bar linkage mechanisms which operate with two independent contact points. The proposed device can deliver a wide range of tactile sensations, such as contact, pressure, twist stimuli, and slippage.

\section{Design of Haptic Display LinkRing}

LinkRing is designed to deliver multi-modal and multi-contact stimuli at the finger pads. The proposed device is based on the planar parallel mechanism, which was previously applied in several works \cite{Tsetserukou2014}, \cite{Altamirano2019}, \cite{moriyama2018development}. Two parallel inverted five-bar mechanisms with 2-DoF each deliver tactile feedback in two independent points on the finger pads. Each of them can generate normal and shear forces at the contact point. And in combination, they can simulate different sizes of grasped objects, the feeling of spinning objects, and the sense of sliding a finger on the surface to the left, to the right, and around the axis perpendicular to the finger pad plane. The prototype consists of 3D-printed parts, namely, motor holders fastened to the finger and links made of flexible PLA and PLA, respectively, and metal spring spacers with diameter 6 mm connecting the system of inverted five-bar linkages (Fig.\ref{fig:main}(b)). The distance between the end-effectors is 26 mm, and it can vary by choosing the different lengths of the spacers. The fastening for the finger was designed as a snap ring to be suitable for different sizes. The specification of the device is shown in Table \ref{spec}.
\begin{table}[h]
\vspace{-0.5em}
\caption{Technical specification  of LinkRing.}
\label{spec}
\centering
\begin{tabular}{lc}
\hline
Motors & Hitec HS-40\\
Material of the motor holders & Soft PLA\\
Material of the links & PLA\\
Weight [g] &  $33$\\
Link length L1, L2 [mm] & $35, 17$\\
Max. normal force at & \\
 each contact point [N] & $1.5 \pm 0.15$
\end{tabular}
\vspace{-1.5em}
\end{table}

\begin{figure}[h]
\centering
\vspace{-0.5em}
\includegraphics[width=0.8\linewidth]{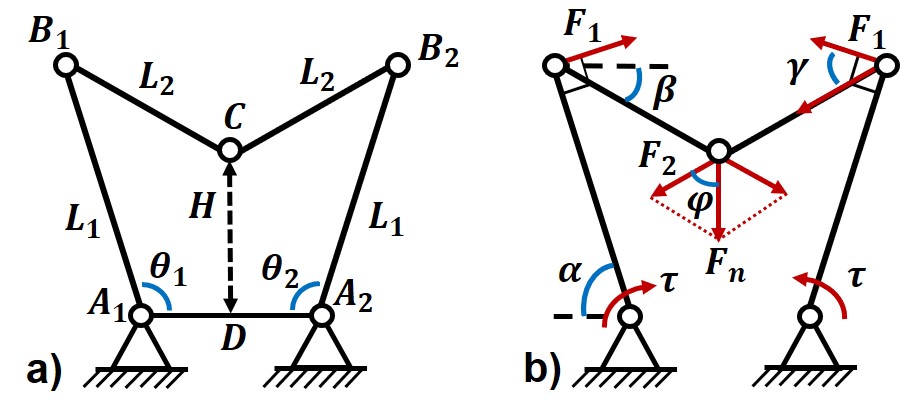}
\qquad
\vspace{-1.5em}
\caption{a) Device kinematic scheme. b) Static force diagram.}
\label{fig:Force}
\vspace{-1.5em}
\end{figure}

The scheme of the wearable haptic display is shown in Fig. \ref{fig:Force}(a). The mechanism has two input links with length $L_1$ controlled by the input angles $\theta_1$, $\theta_2$, two output links with length $L_2$, and the ground link with length D. The developed prototype has the following parameters: $L_1 = 35\ mm$, $L_2 = 17\ mm$, and $D = 15\ mm$. The initial position for the end effector, when the mechanism is symmetric and in contact state with the user's finger, is $H = 22\ mm$. For this position, the input angle is equal to $\alpha = 180-\theta_1=84^{\circ}$, and the angle between the output link and horizontal is the following $\beta = arcsin(\frac{L_1sin\alpha-H}{L_2})=49^{\circ}$. From Fig. \ref{fig:Force}(b) we can calculate the angle $\gamma$ between the force $F_{1}$ and the output link and the angle $\varphi$ between the force $F_{2}$ and the vertical in the following way: $\varphi = 90^{\circ} - \beta=41^{\circ}$ and $\gamma = 90^{\circ} -\alpha + \beta=55^{\circ}$. 

The device is actuated by Hitec HS-40 servo motors with a low mass and rather high output torque (Weight is 4.8 g, dimensions are 20 $\times$ 8.6 $\times$ 17 mm, maximum torque is 0.6 kg-cm). The maximum force is achieved with the symmetric case when the resulting force applied to the finger is a normal force (directed vertically). In this case: $F_1 = \tau /L_1$; $F_2 = F_1\cdot cos(\gamma)$; $F_n = 2\cdot F_2\cdot cos(\varphi)$. The maximum value of the normal force is $F_n = 1.46\ N$. To verify this result, we experimentally measured the force with a calibrated force sensing resistor (FSR 400). The experiment showed that the maximum generated force is $F = 1.5\ N \pm 0.15\ N$.

\section{User study}
\begin{figure}[h]
\centering
\vspace{-0.5em}
\includegraphics[width=\linewidth]{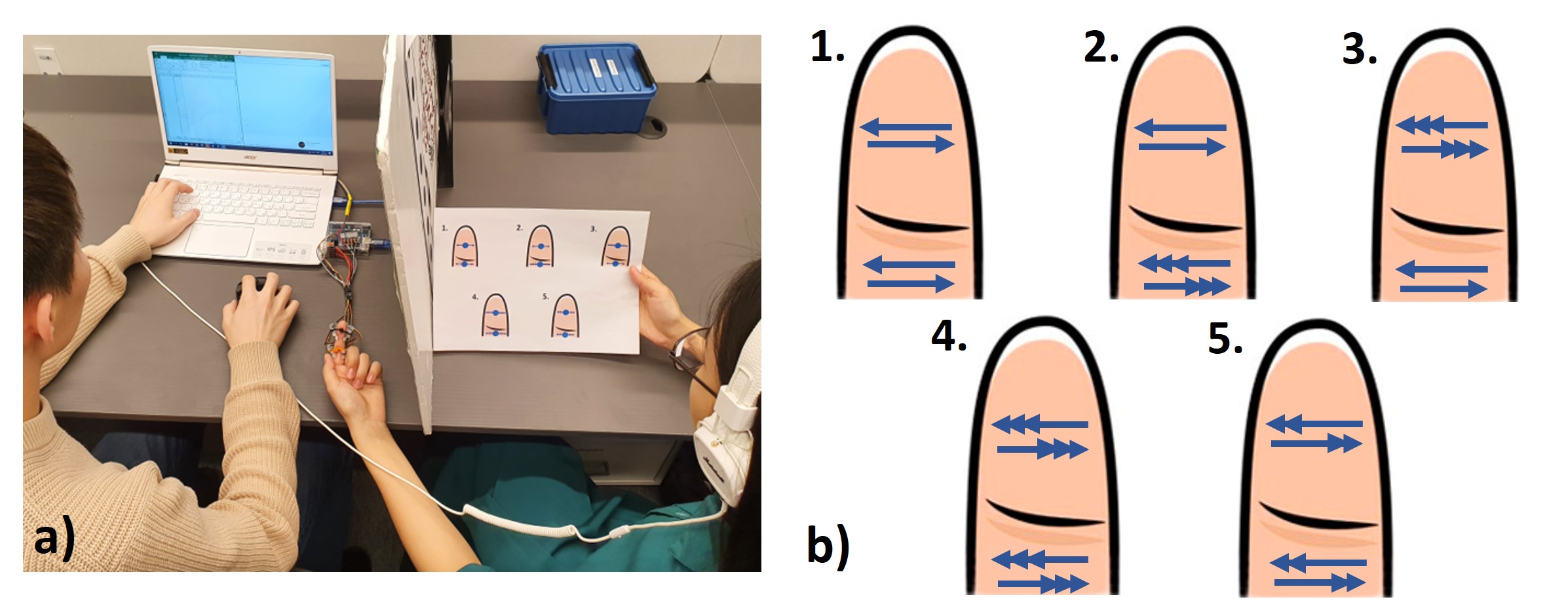}
\qquad
\vspace{-1.5em}
\caption{a) Overview of the experiment. b) Slippage patterns for the experiment. Single arrows indicate the slow speed in the contact point, double arrows mean middle rate, and triple arrows represent fast speed. The direction of the arrow indicates the side of the contact point motion.}
\label{fig:user study}
\vspace{-1.5em}
\end{figure}

We evaluated the performance of the developed haptic display in two user studies. Firstly, we estimated the distinction in the perception of different static contact patterns simulated on the user’s finger.  Secondly,  we tested the ability of users to distinguish the sliding speed of end effectors at the contact area. For the two experiments, the user was asked to sit in front of a desk, and to wear the  LinkRing display on the left index finger. To increase the purity of the experiment, we asked the users to use headphones and fenced off the hand with a device from the user by a barrier (Fig. \ref{fig:user study} (a)).

\subsection{Static pattern experiment}

\begin{figure}[h]
\centering
\includegraphics[width=0.95\linewidth]{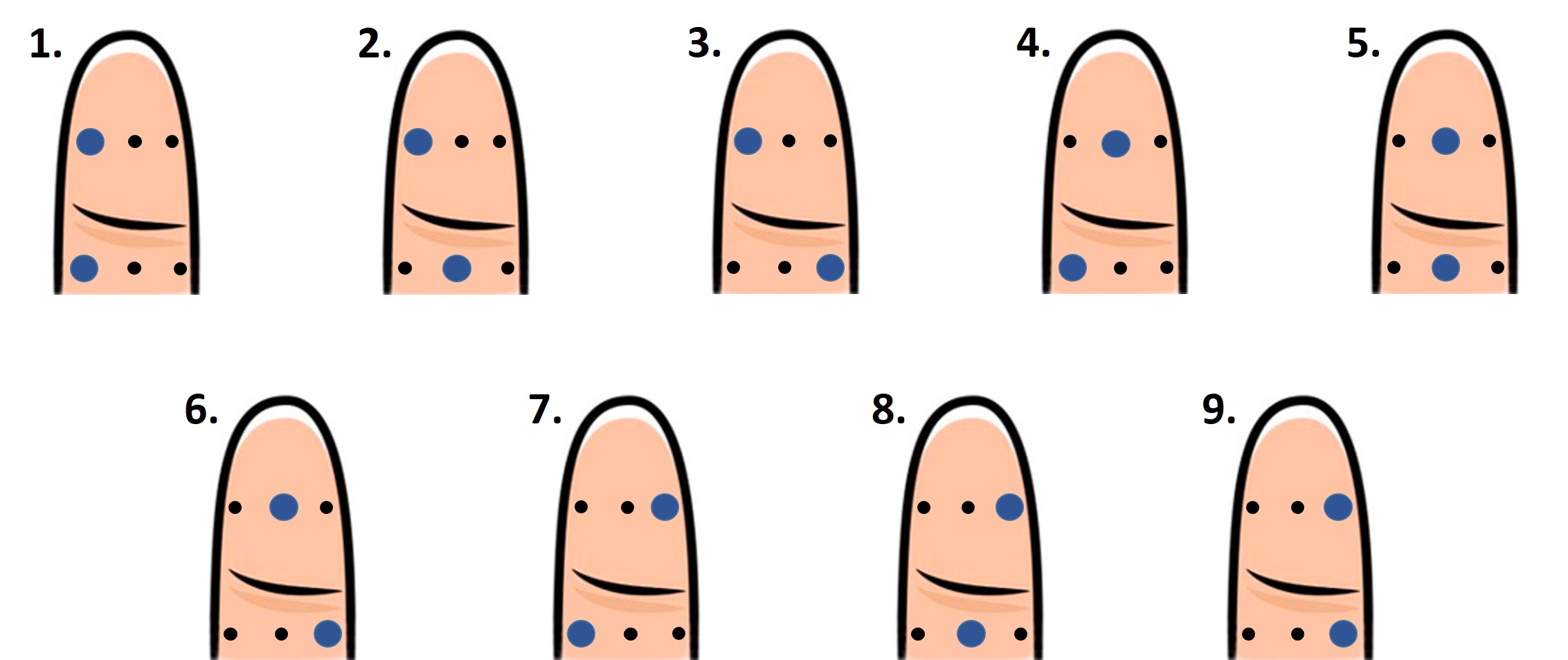}
\qquad
\vspace{-1.0em}
\caption{Patterns for the experiment with static points. Blue dots represent pressed point, and black dots show the possible contact locations.}
\label{fig:static patterns}
\vspace{-0.5em}
\end{figure}

The purpose of the experiment was to study the recognition of various patterns on the user’s finger differing in the location of the contact points. Nine patterns with different positions of static points were designed (Fig. \ref{fig:static patterns}). In total, ten subjects took part in the experiment, two women and eight men, from 21 to 30 years old.

Before the experiment, the calibration was conducted for each participant to provide all the contact points on the finger. The calibration was based on two parameters: the thickness and width of the finger. During the training session, each pattern was delivered two times to the user.

During the experiment, the participant was provided with a visual guide of the designed patterns. For each trial, end effectors reached a specified location in a non-contact position. After that, one pattern was delivered to the finger for three seconds, and they returned to the non-contact position. After the delivery of the static pattern on the user’s finger, the subject was asked to specify the number that corresponds to the provided contact pattern.
Each pattern was delivered five times in random order. In total, 45 patterns were presented to each subject.
\setlength{\parskip}{12pt} \par \noindent \textbf{Experimental results} \setlength{\parskip}{5pt} \par \noindent
Table \ref{cm1} shows a confusion matrix for actual and perceived static patterns. Every row in the confusion matrix represents all 50 times a contact pattern was provided. The results of the experiment revealed that the mean percent of correct answers for each pattern averaged over all the participants ranged from 68$\%$ to 100$\%$. The mean percentage of the correct answers is $90\%$. The most recognizable contact positions were 1st, 4th, and 5th, with a recognition rate of $98\%$, $98\%$, and $100\%$. And the least distinct pattern was 9th with a rate of $68\%$. It can be observed that a high number of participants confused pattern 9 with 8, which are very close. 
\begin{table}[h]
\caption{Confusion matrix for actual and perceived static patterns across all the subjects}
\vspace{-1.5em}
\label{cm1}
\centering
\includegraphics[width=0.8\linewidth]{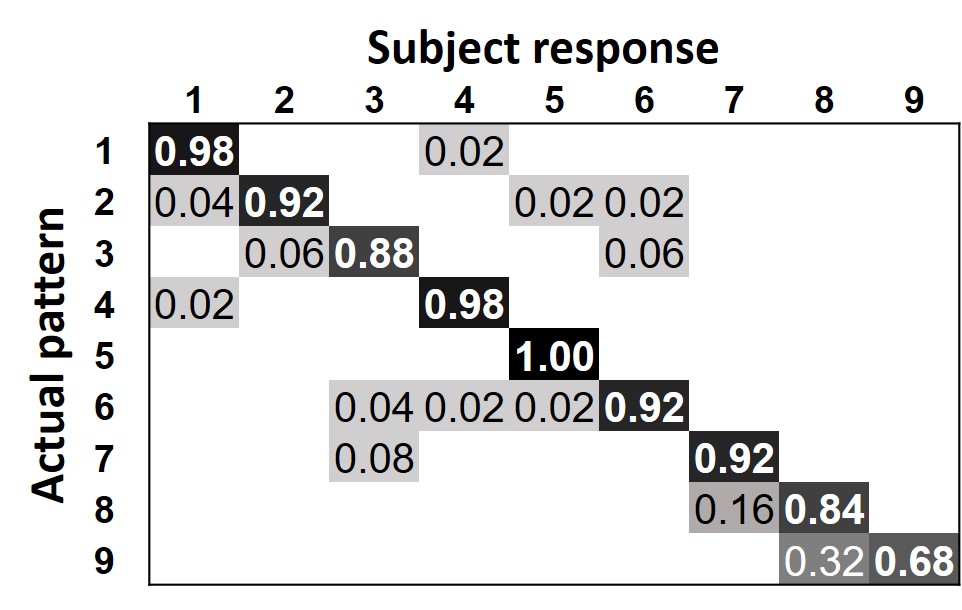}
\vspace{-1.5em}
\end{table}

In order to understand if there is a real difference between pattern perception, the experimental results were analyzed using one-factor ANOVA without replication with a chosen significance level of $p<0.05$. According to the test findings, there is a statistically significant difference in the recognition rates for the different contact patterns $(F(8,81) = 2.43,\ p=0.02<0.05)$. It was significantly more difficult for participants to recognize pattern $9th$ than $5th(F(1,18)  =  16,\ p=8.4\cdot 10^{-4}<0.05)$, $1st$ and $4th(F(1,18)  =13.2,\ p=1.88\cdot 10^{-3}<0.05)$, and $2nd$, $6th$ and $7th(F(1,18)  =  6.23,\ p=0.022<0.05)$.
 
\subsection{Experiment of the recognition of moving patterns} 
The objective of the experiment was to study the user recognition of end effector slippage with various speeds which were delivered to the finger at the same time by both contact points. Eleven participants volunteered into the experiments, two females and nine males, from 21 to 31 years old. 

Three different speeds for slippage on the finger were used: slow (43 mm/s), middle (60 mm/s), and fast (86 mm/s). The five slippage patterns were designed to study the sensing of the end effectors velocity. Second and third patterns transmit different speeds on contact points, and other patterns deliver equal rates (Fig. \ref{fig:user study} (b)). Each pattern was presented five times in random order, thus, 25 patters were provided to each subject.

\setlength{\parskip}{12pt} \par \noindent \textbf{Experimental results} \setlength{\parskip}{5pt} \par \noindent
The results of the experiment are summarized in a confusion matrix (Table \ref{cm2}). The mean percentage of the correct answers is 81$\%$. The most distinctive speed patterns are 2nd and 3rd, with a recognition rate of $94\%$ and $89\%$. They have represented patterns with different velocities of the end effector in two contact points. And the most confusing pattern is 5th with a recognition rate of 65$\%$.

Using the one-factor ANOVA without replications, with a chosen significance level of $p <0.05$, we found a statistically significant difference among the different dynamic patterns $(F(5,50)=3.28,\ p= 1.83\cdot 10^{-2}<0.05)$. According to ANOVA  results, $2nd$ pattern has a significantly higher recognition rate  than the  patterns  $1st(F(1,20)  =  5.75,\ p=0.026<0.05)$ and  $5th(F(1,20)  =  11.4,\ p=  2.97\cdot 10^{-3}<0.05)$. It was significantly easier for participants to recognize the $3rd$ pattern than the $5th(F(1,20)  =  6.17,\ p=0.022<0.05)$.

\begin{table}[h]
\vspace{-0.5em}
\caption{Confusion matrix for actual and perceived slippage patterns across all the subjects}
\vspace{-1.5em}
\label{cm2}
\centering
\includegraphics[width=0.7\linewidth]{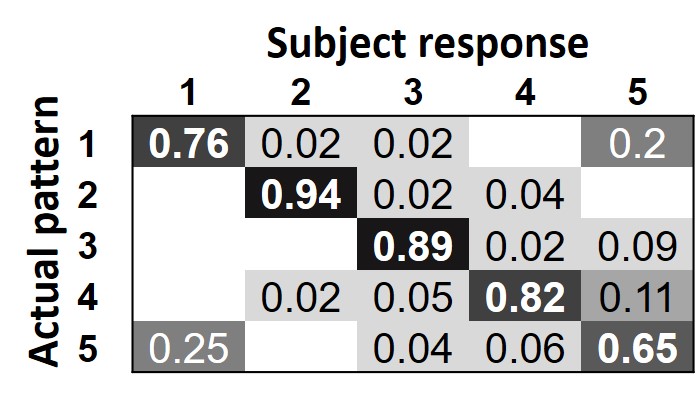}
\vspace{-1.5em}
\end{table}

\section{Conclusion}

We have developed LinkRing, a wearable haptic display that can provide multi-contact stimuli in two independent points of the user’s finger. The device is capable of generating a wide range of tactile sensations such as contact, slippage, twist stimuli, and pressure. The structure of the device is lightweight and easy to wear. The user study revealed high recognition rates in discrimination of static and dynamic patterns delivered to the finger pads. The obtained results allow us to determine the most suitable patterns for further presenting the static and moving object for the finger perception with the proposed display.

The future work will be aimed at expending multi-modal stimuli by adding vibration motors to the end effectors as well as improving the design of the device by reducing its dimensions and increasing its ergonomics. Various virtual applications are going to be developed to study virtual immersion quality and fidelity of multi-modal tactile stimuli. The developed haptic display can potentially bring a highly immersive VR experience in the guiding blind navigation systems, teleoperation, and medical VR simulators.

\bibliographystyle{splncs04}
\bibliography{bib}

\end{document}